\documentclass[eqnumis, pdf]{hjms}

\usepackage{amsfonts}
\usepackage{amsmath}
\usepackage{footnote}
\usepackage{multicol}
\usepackage{booktabs}
\usepackage[flushleft]{threeparttable}
\usepackage[font=small,labelfont=bf]{caption}

\usepackage{hyperref}
\hypersetup{
colorlinks=true,
urlcolor=blue,
citecolor=blue}

\newcommand{\defn}{\stackrel{\Delta}{=}}

\begin{document}
%

\markboth{H. Af\c{s}er}{Some Remarks on Bayesian Multiple Hypothesis Testing}

\title{Some Remarks on Bayesian Multiple Hypothesis Testing}

\author{H\"{u}seyin Af\c{s}er}

\address{Department of Electrical-Electronics Engineering, Adana Alparslan T\"{u}rke\c{s} \\ Science and Technology University, Adana, Turkey}
\emails{afser@atu.edu.tr}
\maketitle
\begin{abstract}

	We consider Bayesian multiple hypothesis problem with independent and identically distributed observations. The classical, Sanov's theorem-based, analysis of the error probability allows one to characterize the best achievable error exponent. However, this analysis does not generalize to the case where the true distributions of the hypothesis are not exact or partially known via some nominal distributions. This problem has practical significance, because the nominal distributions may be quantized versions of the true distributions in a hardware implementation, or they may be estimates of the true distributions obtained from labeled training sequences as in statistical classification. In this paper, we develop a type-based analysis to investigate Bayesian multiple hypothesis testing problem. Our analysis allows 
one to explicitly calculate the error exponent of a given type and extends the classical analysis. As a generalization of the proposed method, we derive a robust test and obtain its error exponent for the case where the hypothesis distributions are not known but there exist nominal distribution that are close to true distributions in variational distance.  
	
%

\end{abstract}



\section{Introduction}
\label{Int}

In the last decades, multiple hypothesis testing (MHT) framework found diverse usage in different fields of engineering. For example, in  \cite{Trachi} authors use MHT to design a machine fault detector where the unknown model parameters  are estimated from the available data, then MHT is used for detection. In a similar fashion, authors in \cite{Aizpurua} used MHT for
fault detection in power transformers. In \cite{Korki} MHT is used to detect and recover block-sparse signals with unknown block structures. In \cite{Makris} authors applied MHT to multiple-target tracking problem via a two-step MHT tracking algorithm where the model state is estimated first, then the tracing is accomplished with the MHT. The authors in \cite{Lee} investigated the energy detection problem with mismatched distributions  and characterized its asymptotic performance via large deviations theory. In \cite{Ali} MHT is used to provide frame synchronization in multiple access communication channels.

In this paper we consider Bayesian MHT where the average error probability is a weighted sum of the error probabilities under each hypothesis and hypotheses are assumed to have positive prior probabilities. This setup differs from Neyman-Pearson type MHT where the aim is to minimize the error probability under a selected hypothesis by limiting the remaining error probabilities to specified values \cite{Tuncel}.


The asymptotic performance analysis of Bayesian MHT is Based on Sanov's theorem and large deviations theory \cite[Ch. 11.4]{Cover}. This analysis reveals that the exponential speed with which the error probability decays equals to the minimum pairwise Chernoff information between the distinct hypothesis distributions. This achievability result is originally formalized by Leang et al. \cite{Leang}. Then, Westover provided a Sanov's theorem-based proof for the asymptotic geometry of Bayesian MHT \cite{Westover}. The asymptotic analysis often ignores the effect of the prior probabilities of hypothesis as they vanish as the length of the test sequence gets large. Moulin \cite{Moulin} investigated asymptotic performance of Bayesian and Neyman-Pearson MHT, by taking into account the effect of priors, and refined the asymptotic bounds in \cite{Leang, Tuncel} for the case where the length of the test sequences takes moderate values. The non-asymptotic performance analysis of Bayesian binary hypothesis testing is investigated by Sason in \cite{Sason_2}, and a finite-length analysis of Neyman-Pearson binary hypothesis testing can be found in \cite{Espinoza}.

The classical Sanov's theorem-based analysis can not be used if the hypothesis distributions are not exact or they are partially known via some nominal distributions. This problem has practical significance, because the nominal distributions may be fixed point counterparts of the true distributions in a hardware implementation, or they can be estimates of the true distributions obtained from labeled training sequences as in statistical classification \cite{Hastie,Gutman, Afser}. In this paper we provide a type-based framework for analyzing the asymptotic performance of Bayesian MHT. Our method generalizes Sanov's theorem-based method by allowing one to obtain upper bounds on the error probability of a specific input type.  Then, motivated by classical robust hypothesis testing \cite{HuberB,HuberA,Levy}, we generalize our analysis to the case where the distributions of the hypothesis are not known exactly but one has access to a set of nominal distributions that are close to true ones. For this problem we propose a low-complexity, type-based robust test and obtain an upper bound on its error probability. We show that the proposed test and the upper bound tend to optimal as the uncertainties in distributions vanish. We compare its performance to a well-known robust test, namely the DGL test due to Devroye et al. \cite{Devroye}, and show that it provides performance advantages when the uncertainties in hypothesis distributions are arbitrarily small.

The outline of the paper is as follows: We provide some  preliminary material in Section 2 and present the problem definition. We review the classical Sanov's theorem-based analysis of the Bayesian multiple hypothesis testing and present the proposed type-based analysis in Section 3. In Section 4, we apply the proposed analysis to a robust hypothesis testing problem and present our simulations. Finally,  Section 5 concludes the paper and provides directions for future research.

\section{Preliminaries} \label{section:Sect_2}

\subsection{Notation}
We use upper case letters $X$, $Y$ to denote random variables and lower cases $x$, $y$ for their realizations. We let $\mathcal X$ denote the alphabet of $X$ and use $|\mathcal X|$  to denote the size of this alphabet. We write $P(x)$ to denote the probability of the event $X=x$, and similarly $P_i(x)$ is used to denote the probability of $X=x$ when $X$ is generated from distribution $P_i$. 

\subsection{Statistical distances}
The following, are the definitions of the statistical distances that are used in this paper. Let $P$ and $Q$ be two distributions that are defined over a common and discrete alphabet $ \mathcal{X}$.
The variational distance (or the total variation) between $P$ and $Q$ is defined as
\begin{align}
V(P,Q)= \frac{1}{2}\sum_{x \in \mathcal{X}} |Q(x)-P(x)|. 
\end{align}

KL divergence from  $P$ to $Q$ is calculated as
\begin{align}
D( P || Q ) = \sum_{x \in \mathcal{X}} P(x) \log \frac{P(x)}{Q(x)}. 
\end{align}
where the base of the logarithm is $2$ (this convention is adopted throughout the paper).

Chernoff information (distance) plays a crucial role for the analysis of Bayesian hypothesis testing and is defined as
\begin{align}
C(P,Q) = -\underset{\lambda \in [0,1]}{ \text{min}} \log \left( \sum_{x \in \mathcal{X}} P(x)^\lambda Q(x)^{1-\lambda} \right).
\end{align}

Following is an inequality between Chernoff information and variational distance \cite{Sason}
\begin{align}
C(P,Q) \geq -\frac{1}{2} \ln (1 -V(P,Q)^2) \label{eq:Sason_ineq}
\end{align}
where $\ln$ denotes the natural logarithm.

\subsection{Method of types}

The analysis in this paper makes heavy use of the method of types \cite[Chapter 11]{Cover}. Below, we give some definitions and key findings that are crucial for the upcoming analysis.
\begin{itemize}
\item The type vector $P_{{\vec{x}}}$:  It is the empirical probability distribution of the vector ${\vec{x}}$, ${\vec{x}} \in {\mathcal X}^n$.
\begin{align}
P_{{\vec{x}}}(a) \defn \frac{1}{n}N(a|{\vec{x}}), \quad \quad \forall a \in {\mathcal{X}}
\end{align}
where $N(a|{\vec{x}})$ denotes the number of occurrences of symbol $a$ in vector ${\vec{x}}$. Notice that $P_{{\vec{x}}}$ is a proper distribution for the alphabet $\mathcal X$.

\item The type class $T(P)$: It  is the set of all sequences ${\vec{x}}$, ${\vec{x}} \in {\mathcal X}^n$, that have type $P$ i.e. $T(P) = \{ {\vec{x}} : P_{{\vec{x}}}=P \}$.

\item The set of all type classes ${\mathcal{P}}^n$:  For fixed $n$, this set is defined as  $ {\mathcal{P}}^n = \{ T(P) : T(P)  \neq \emptyset \}$.

\end{itemize}

\begin{lemma}\label{thm:lemma_1}
Size of the type class $T(P)$ satisfies
\begin{align}
(n+1)^{-|{\mathcal{X}}|}2^{nH(P)}\leq| T(P) | \leq2^{nH(P)}.
\end{align}
\end{lemma}
Here, $H(P)$ denotes the entropy of the distribution $P$
\begin{align}
 H(P) = - \sum_{x \in \mathcal{X}} P(x) \log P(x).
\end{align}
\begin{lemma}\label{thm:lemma_2}
The  total number of type classes is upper bounded as
\begin{align}
|{\mathcal{P}}^n |\leq (n+1)^{|{\mathcal{X}}|-1}.
\end{align}
\end{lemma}

\begin{lemma}\label{thm:lemma_3}
The vectors of the same type has the same probability. If the elements of ${\vec{x}}$ are generated independently from distribution $P$, then 
\begin{align}
P({\vec{x}}) = 2^{-n \left( H(P_{{\vec{x}}}) + D(P_{{\vec{x}}}|| P) \right)}.
\end{align}
\end{lemma}

\subsection{Problem Statement}
Bayesian MHT addresses the following problem: For an observed vector $\vec{X}=[X_1,X_2,...\\
.,X_n] \in {\mathcal X}^n$ decide on one of $M$ hypothesis ${\mathcal H}_1,{\mathcal H}_2,....,{\mathcal H}_M$ with distributions $P_1,P_2,....,P_M$ where ${\mathcal H}_i$, $i=1,2,...,M$, states that the realization $\vec{X}=\vec{x}$ is independent and identically distributed according to $P_i$ \cite{Lehman}. In this paper, we assume that the distributions are defined over a common alphabet ${\mathcal X}$ which is discrete and countably finite. In order to perform the test, one seeks a decision rule that partitions ${\mathcal X}^n$ into $M$ mutually exclusive and collectively exhaustive acceptance regions $\Omega_1,\Omega_2,....,\Omega_M$ such that ${\mathcal H}_i$ is accepted if $\vec{x} \in \Omega_i$. Let $P(e|{\mathcal H}_i)$ denote the probability of error when ${\mathcal H}_i$ is true but the test
decides otherwise. In the Bayesian setting, one assumes strictly positive priors $P({\mathcal H}_1),P({\mathcal H}_2),....,P({\mathcal H}_M)$ for the $M$ hypothesis and the resultant probability of
error equals to
\begin{align}
P(e) = \sum_{i=1}^M P(e|{\mathcal H}_i) P({\mathcal H}_i).
\end{align}
The minimization of the above probability is accomplished with maximum a posteriori (MAP) decision rule.  When $n$ is sufficiently large, the effect of the priors vanishes and MAP
decision rule simplifies to the nearest neighbor (NN) decision rule  as
\begin{align}
\text{Choose ${\mathcal H}_i$},  \quad i = \underset{j \in \{1,2,...,M\}}{\text{argmin}}{D(P_{{\vec{x}}}| P_j)} \label{eq:NN_rule}.
\end{align}
The test partitions the probability simplex into $M$ disjoint regions where $P_j$, $j=1,2,....,M,$ are the centroids. The decision rule measures the KL divergence
(distance) between $P_{\vec{x}}$ and $P_j$ and assigns $\vec{x}$ to the nearest region.

\section{Error probability of Bayesian MHT} \label{section:Sect_2}

The classical analysis of the error probability of Bayesian MHT is based on Sanov's theorem \cite[p. 362]{Cover} which allows one
to analyze the probability of having $\vec{x} \in \Omega_i$, $i=1,2,...,M$, under different hypothesis. The result of this analysis states that the \textit{best} achievable error
exponent of binary Bayesian hypothesis testing equals to the minimum Chernoff information of hypothesis distributions \cite[Theroem 11.9.1]{Cover}. However, this analysis
is not explicit in the sense that the average error probability of a specific type i.e. $T(P_{\vec{x}})$ can not be calculated. This flexibility may be important in Bayesian hypothesis testing with a rejection option where the test is allowed to make a no-match decision as in \cite{Gutman}. Because, by rejecting the type classes with smallest error exponents the performance of the test can be improved. 

In this section, we propose a type-based analysis that does not require Sanov's theorem and allows one to upper bound the error probability  averaged over given type class $T(P_{\vec{x}})$. The proposed analysis also reveals that the \textit{worst} case achievable error exponent over the distinct types equals to the minimum Chernoff information, therefore this exponent is always achievable.

First, we are going to explain the classical Sanov's theorem-based analysis in \cite{Westover}. Then, we are going the present the proposed analysis and make connections between them.

\subsection{Classical analysis}
 
Considering the NN decision rule in \eqref{eq:NN_rule}, with acceptance regions $\Omega_1,\Omega_2,....,\Omega_M$, let $\Omega_i^c$ denote the complement of $\Omega_i$.
The application of Sanov's theorem gives 
\begin{align}
P(e|{\mathcal H}_i) & \leq 2^{(-nD(P_i*|P_i) - \frac{(|{\mathcal X}|-1) \log (n+1)}{n} )}, \\
P_i^* & \defn \underset{ {\bf p} \in \Omega_i^c}{\text{argmin}} \quad {D( {\bf p}| P_i)},
\end{align}
Using the above result, one calculates the total error probability as
\begin{align}
P(e) &= \sum_{i=1}^M P(e|{\mathcal H}_i)P({\mathcal H}_i) \nonumber \\
& \leq \sum_{i=1}^M 2^{(-nD(P_i^*|P_i) - \frac{(|{\mathcal X}|-1) \log (n+1)}{n} )} \nonumber \\
& \leq M  \max_{i} 2^{(-nD(P_i^*|P_i) - \frac{(|{\mathcal X}|-1)\log (n+1)}{n} )}, \quad i=1,2,...,M  \nonumber \\
& =  2^{-n \left( \min_{i} D(P_i^*|P_i) - \frac{(|{\mathcal X}|-1)\log (n+1)}{n} - \frac{\log M}{n}\right)}, \quad i=1,2,...,M \nonumber \\
& \leq 2^{-n \left( \underset{i \neq j} {\text{min }}C(P_i,P_j) - \frac{|{\mathcal X}-1| \log (n+1)}{n} - \frac{\log M}{n} \right)}, \quad i=1,2,...,M, \quad j=1,2,...,M,  \label{eq:exp_id}
\end{align}
where the last step follows from the fact that the minimum value of $\min_{i} D(P_i^*|P_i)$ equals to $ {\text{min }_{i \neq j}C(P_i,P_j)}$ when $P_i^*$ is of the form
\begin{align}
P_i^* = \frac{ P_i(x)^\lambda P_j(x)^{1-\lambda}}{\sum_{x \in {\mathcal X}}  P_i(x)^\lambda P_j(x)^{1-\lambda} }
\end{align}
and $\lambda$ is chosen such that $ D(P_i^*|P_i)=D(P_i^*|P_j)$ \cite{Westover}.

\subsection{Proposed  analysis}
 
 The general approach to performance analysis of communications systems is to investigate the error probability of the system for a given input. Then, the performance
 of the system can be analyzed by averaging this error probability over all possible inputs. Here, we follow a similar approach and first investigate the probability of error for a given type $P_{\vec{x}}$. In this regard, let us define $P(e |P_{\vec{x}} )$ to be the probability of error, averaged over $\vec{x} \in T(P_{\vec{x}})$. 
\begin{align}
P(e | P_{\vec{x}} ) \defn \sum_{\vec{x} \in T(P_{\vec{x}})} P(e| \vec{x})P(\vec{x}).
\end{align}
Then, the probability of error averaged over $\vec{x} \in {\mathcal X}^n$ can be calculated as 
\begin{align}
P(e) &= \sum_{ \vec{x} \in {\mathcal X }^n } P(e| \vec{x})P(\vec{x}) \\
	        &= \sum_{  T(P_{\vec{x}}) \in {\mathcal P }^n } \sum_{ \vec{x} \in T(P_{\vec{x}}) } P(e|  \vec{x})P(\vec{x}) \\
	        &= \sum_{  T(P_{\vec{x}}) \in {\mathcal P }^n } P(e | P_{\vec{x}} ).  \label{eq:exp_sum}
\end{align}
If $P(e | P_{\vec{x}} )$ terms in \eqref{eq:exp_sum} are exponentially decaying with certain exponents that depend on $P_{\vec{x}}$, then the minimum of them must dominate the exponent of $P(e)$. The following propositions, whose proof are provided in the Appendix, show that this is infact the case. 
\begin{proposition}\label{thm:prop_1}
\begin{align}
P(e |P_{\vec{x}} )  \leq 2^{-n \left(  \underset{i \neq j}{\min} \hspace{3pt} \max \{ D(P_{{\vec{x}}}||P_i), D(P_{{\vec{x}}}||P_j) \} - \frac{\log M}{n} \right)},  \quad i=1,2,...,M, \quad j=1,2,...,M.
\end{align}
\end{proposition}
\begin{proposition}\label{thm:prop_2}
\begin{align}
\underset{T(P_{\vec{x}}) \in {\mathcal P }^n}{\min } \hspace{3pt}  \max \{ D(P_{{\vec{x}}}||P_i), D(P_{{\vec{x}}}||P_j) \} = C(P_i,P_j),
\end{align}
and the minimizing $P_{\vec{x}}$ is of the form
\begin{align}
P_{\vec{x}}^\lambda = \frac{ P_i(x)^\lambda P_j(x)^{1-\lambda}}{\sum_{x \in {\mathcal X}}  P_i(x)^\lambda P_j(x)^{1-\lambda} } \label{eq:min_type}
\end{align}
where $\lambda$ is chosen such that $ D(P_{\vec{x}}^\lambda|P_i)=D(P_{\vec{x}}^\lambda|P_j)$.
\end{proposition}

Proposition 1 indicates that $P(e |P_{\vec{x}} )$ is decaying with an exponent $ \underset{i \neq j}{\min} \hspace{3pt}  \max \{ D(P_{{\vec{x}}}||P_i),$ \newline
$D(P_{{\vec{x}}}||P_j) \}$, and Proposition 2 implies that the minimum of this exponent over distinct types equals to $\underset{i \neq j}{\min} \hspace{3pt}  C(P_i,P_j)$. This fact is demonstrated with a simple  example in Figure 1 where
we have plotted  the ratio $ \underset{i \neq j}{\min} \hspace{3pt}  \max \{ D(P_{{\vec{x}}}||P_i), D(P_{{\vec{x}}}||P_j) \} / \underset{i \neq j}{\min} \hspace{3pt}  C(P_i,P_j)$
by sorting it  for all $T(P_{\vec{x}}) \in {\mathcal P }^n$. As can be seen, the minimum value of this ratio equals to $1$ whereas it can be significantly larger depending on  $T(P_{\vec{x}})$.

Finally, the probability of error averaged over $\vec{x} \in {\mathcal X}^n$ can be upper bounded as follows. From \eqref{eq:exp_sum} we obtain
\begin{align}
P(e)     &\leq | {\mathcal P }^n | \max_{ T(P_{\vec{x} })    \in {\mathcal P }^n }  P(e | P_{\vec{x} } ) \label{eq:pe_1}\\
	        &\leq  |{\mathcal P }^n |  2^{-n \left(  \underset{i \neq j}{\min} \hspace{2 pt} \underset{ T(P_{\vec{x} })    \in {\mathcal P }^n }{\min}    \hspace{3pt} \max \{ D(P_{{\vec{x}}}||P_i), D(P_{{\vec{x}}}||P_j) \} - \frac{\log M}{n} \right)},   \\
	        &=  |{\mathcal P }^n |  2^{-n \left(  \underset{i \neq j}{\min}  \hspace{2 pt} C(P_i,P_j) - \frac{\log M}{n} \right)} \\
	        & \leq 2^{-n \left(  \underset{i \neq j}{\min}  \hspace{2 pt} C(P_i,P_j) -  \frac{(|{\mathcal X}|-1) \log (n+1)}{n} - \frac{\log M}{n} \right)}, \quad i=1,2,...,M, \quad j=1,2,...,M.  \label{eq:pe_2}
\end{align}
where the last ineqaulity follows from Lemma 2.

\begin{figure*}[t]
\begin{center}
\includegraphics[scale=0.5]{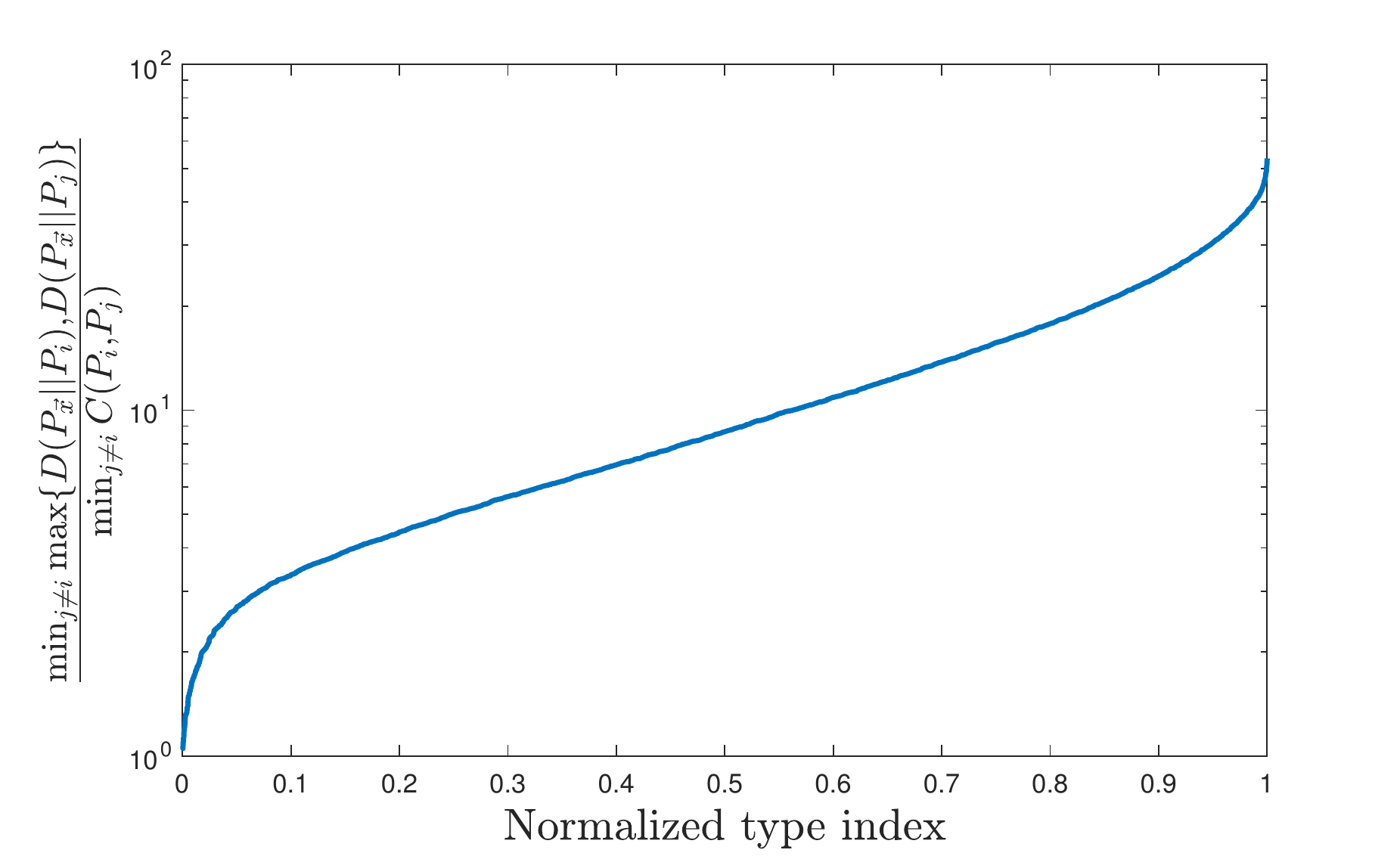}
\caption{ The sorted ratio of the achievable exponents of $P(e|P_{\vec{x}})$ in Proposition 1 and the minimum Chernoff information. In this example, $M=5$, $|{\mathcal X}|=3$, 
 $P_1= [0.1,0.8,0.1],
P_2= [0.3,0.2,0.5], 
P_3= [0.6,0.1,0.3], 
P_4= [0.4,0.4,0.2]$ and 
$P_5= [0.3,0.6,0.1].
$}
\label{fig: fig_1}
\end{center}
\end{figure*}

By comparing the classical analysis and the proposed analysis one can easily observe that the minimization of $\min_{i} D(P_i^*|P_i)$ terms in \eqref{eq:exp_id} corresponds to finding the type class that minimizes $ \max \{ D(P_{{\vec{x}}}||P_i), D(P_{{\vec{x}}}||P_j) \}$ in the proposed method. However, this is not explicit in classical analysis due to the usage of Sanov's theorem. Our analysis exposes this fact with Propositions \ref{thm:prop_1} and \ref{thm:prop_2} and complements the classical analysis. Moreover, Proposition \ref{thm:prop_1}  can be used to obtain an upper bound on the error probability of a specific type $P_{\vec{x}}$, and this bound will be tighter than \eqref{eq:exp_id} if $P_{\vec{x}}$ 
 is different than \eqref{eq:min_type}. 

\section{Application:  a robust hypothesis testing problem} \label{section:Sect_3}

As an application of the proposed analysis, we consider a robust hypothesis testing problem where the distributions of the hypothesis are not known exactly but one has access
to a set of nominal distributions $\{ Q_1,Q_2,....,Q_M \}$. Here, the nominal distributions are close to actual distributions in variational distance as 
\begin{align}
V(P_j, Q_j) \leq \epsilon_j,  \quad j=1,2,...,M \label{eq:dist_const}
\end{align}
where $\epsilon_j$ are known robustness parameters.

The above problem, in general, is studied within the context of classical robust hypothesis testing where the alphabet ${\mathcal X}$ can be discrete or continuous i.e. $|\mathcal X|=\infty$. In this setting, the usual procedure is 
to follow a minimax approach and to minimize the worst case probability of error under the distance constraint in \eqref{eq:dist_const} \cite{HuberB,HuberA,Levy}. However, this approach can not be  extended to multiple hypothesis and is explicit only when $M=2$. For multiple hypothesis, the closest work to the one considered here  is the DGL
test due to Devroye et al. \cite{Devroye}, but as reported in \cite{Biglieri} this test can be strictly sub-optimal as the uncertainties in the distributions vanish i.e. as $Q_j \rightarrow P_j$, $j=1,2,...,M$. Our approach can be regarded as a generalization of the classical hypothesis testing problem for the case $V(P_j, Q_j) \leq \epsilon_j$, in the discrete setting and we show that the proposed test does not compromise from optimality in the sense that it coincides with the optimal test as $Q_j \rightarrow P_j$. Furthermore,
when $|\mathcal X|$ is finite the achievable exponent of the proposed test is larger than that of  DGL test in the regime $ \epsilon_j | {\mathcal X}| \ll 1$.

The proposed method is based on a rounding operation of the nominal 
distributions to obtain representatives for actual distributions, then using them in the classical hypothesis test. The representative distributions $\bar{P}_1,\bar{P}_2,....,\bar{P}_M$
are obtained respectively from $Q_1,Q_2,...,Q_M$ via the following transformation
\begin{align}
\bar{P}_j(x) = \frac{Q_j(x)+\epsilon_j}{1+|{\mathcal X}|\epsilon_j}, \quad \forall x \in \mathcal X.
\label{eq:p_bar_defn}
\end{align}

It is clear that $\bar{P}_j$ is a proper distribution over the alphabet $\mathcal{X}$ since
\begin{align*}
\sum_{x \in\mathcal{X}} \bar{P_j}(x)&=\frac{\sum_{x \in \mathcal{X}}(Q_j(x)+\epsilon_1)}{1+|{\mathcal X}|\epsilon_1} = \frac{1+|{\mathcal X}|\epsilon_1}{1+|{\mathcal X}|\epsilon_1} = 1
\end{align*}
and $P_j(x) \leq 1$ holds $\forall x \in \mathcal X$. Let us define $U$ as the uniform distribution for the alphabet $\mathcal{X}$
\begin{align}
U(x) \triangleq 
\begin{cases}  \frac{1}{| \mathcal{X} |},  \quad x \in \mathcal{X}, \\
0, \quad  \text{otherwise}.   
\end{cases}
\end{align}
In the light of the above, one can view $\bar{P}_j(x)$ as a rounded version of $Q_j(x)$ where the rounding is towards $U$ and it increases as $\epsilon_j$ increases. To see this, note that the condition $\bar{P}_j(x)  > Q_j(x)$ is equivalent to
\begin{align*}
\frac{Q_j(x)+\epsilon_j}{1+|{\mathcal X}|\epsilon_j}  &> Q_j(x) \\
Q_j(x)+\epsilon_j  &> Q_j(x)+Q_j(x)|{\mathcal X}|\epsilon_j \\
  \frac{1}{|{\mathcal X}| }&> Q_j(x). 
\end{align*}
Thus, if $ Q_j(x)<\frac{1}{|{\mathcal X}| }$ then $\bar{P}_j(x)  > Q_j(x)$ and similarly  $Q_j(x) > \frac{1}{|{\mathcal X}| }$ indicates $\bar{P}_j(x)  < Q_j(x)$. Consequently, and from \eqref{eq:p_bar_defn}, we observe that as $\epsilon_j$ increases
$\bar{P}_j(x)$ gets away from $Q_j$ and closer to $U$ because
\begin{align}
\lim_{\epsilon_j \rightarrow \infty} \bar{P}_j(x) = \lim_{\epsilon_j \rightarrow \infty} \frac{Q_j(x)+\epsilon_j}{1+|{\mathcal X}|\epsilon_j} = U(x).
\end{align}
On the other hand, as $\epsilon_j$ decreases $\bar{P}_j(x) $ gets closer to $Q_j(x)$ and the distance constraint $V(P_j,Q_j) \leq \epsilon_j$ indicates that $Q_j(x)$  gets closer to $P_j$. This, in turn, implies that $\bar{P}_j(x)$ gets closer to $P_j(x)$ and  
\begin{align}
\lim_{\epsilon_j \rightarrow 0} \bar{P}_j(x) = \lim_{\epsilon_j \rightarrow 0} \frac{Q_j(x)+\epsilon_j}{1+|{\mathcal X}|\epsilon_j} = Q_j(x)=P_j(x). \label{eq:P_bar_eq}
\end{align}

The main use of the distribution $ \bar{P}_j$ is to provide an upper bound on $P({\vec{x}})$ when ${\vec{x}}$ is generated according to $P_j$. This fact is provided in the following proposition whose proof is provided in the Appendix.
\begin{proposition}\label{thm:prop_3}
$ \forall {\vec{x}} \in T(P_{\vec{x}})$, given  $V(P_j,Q_j) \leq \epsilon_j$, and independently of $P_j$
\begin{align}
P_j({\vec{x}}) \leq 2^{-n \left( H(P_{{\vec{x}}})+D(P_{{\vec{x}}}||\bar{P}_j)-\log(1+|{ \mathcal X}|\epsilon_j) \right)}. \label{eq:prop_1}
\end{align}
\end{proposition}

Proposition \ref{thm:prop_3}  can be regarded as a generalization of Lemma \ref{thm:lemma_3}  when the true distribution $P_j$ that generated ${\vec{x}}$ is not known exactly, but one has the knowledge that $V(P_j,Q_j) \leq \epsilon_j$. Notice that as $\epsilon_j \rightarrow 0$ the upper bound in Proposition \ref{thm:prop_3} matches the equality in Lemma \ref{thm:lemma_3}.

The proposed test and the upper bound on its error probability is presented in the following theorem. 
\begin{theorem}\label{thm:thm_1}
For the robust Bayesian MHT problem, the total error probability of the decision rule
\begin{align}
\text{Choose ${\mathcal H}_j$},  \quad j = \underset{i \in \{1,2,...,M\}}{\text{argmin}}{D(P_{{\vec{x}}}| \bar{P}_i)} \label{eq:prop_test}
\end{align}
is upper bounded as
\begin{align}
P(e) &\leq 2^{-n \left( \underset{i \neq j}{\min }C(\bar{P}_i,\bar{P}_j)- \log(1+|{ \mathcal X}|\epsilon)-  \frac{(|{\mathcal X}|-1) \log (n+1)}{n} - \frac{\log M}{n} \right)}, \quad i,j=1,2,...,M. \label{eq:exp_bound}
\end{align}
where
\begin{align}
\epsilon \defn \max_{k} \epsilon_k, \quad k=1,2,....,M.
\end{align}

\end{theorem}

\begin{proof}
In the proof of Proposition 1, using Proposition \ref{thm:prop_3} instead of Lemma \ref{thm:lemma_3} results in 
\begin{align}
P(e |P_{\vec{x}} )  \leq 2^{-n \left(  \underset{j \neq i}{\min} \hspace{3pt} \max \{ D(P_{{\vec{x}}}||\bar{P}_i), D(P_{{\vec{x}}}||\bar{P}_j) \} - \log(1+|{ \mathcal X}|\epsilon)- \frac{\log M}{n} \right)}.
\end{align}

Then, the proof follows from Proposition 2 and \eqref{eq:pe_1}-\eqref{eq:pe_2} by using $\bar{P}_j$ instead of $P_j$, $j=1,2,...,M$.
\end{proof}
The implications of Theorem \ref{thm:thm_1} are summarized below.

\vspace{5 pt}
\textit{i)} By comparing  \eqref{eq:NN_rule} and  \eqref{eq:prop_test}, we observe that the proposed test is identical in form to the NN test. The only distinction is that in the proposed test  $\bar{P}_1,\bar{P}_2,...,\bar{P}_M$ must be used instead of the true distributions.  As $\epsilon_j \rightarrow 0$, from \eqref{eq:P_bar_eq} we see that
$\bar{P}_j \rightarrow P_j$, $j=1,2,...,M$, and the proposed test becomes identical to the optimal test. 
 

\vspace{5 pt}
\textit{ii)} By investigating the upper bound in \eqref{eq:exp_bound}, we observe that the proposed test achieves a positive error exponent provided that
\begin{align}
 \underset{i \neq j}{\text{min }}C(\bar{P}_i,\bar{P}_j)- \log(1+|{ \mathcal X}|\epsilon)>0, \quad i=1,2,...,M, \quad j=1,2,...,M.
\end{align}
The above condition can be easily checked prior to performing the test and one can safely use the proposed test if this condition is satisfied. 
In the next section, we show by simulations than even if the above condition is not met the proposed test may still provide an acceptable performance. 
As $\epsilon \rightarrow 0$, we have  $\bar{P}_j \rightarrow P_j$ and $\log(1+|{ \mathcal X}|\epsilon) \rightarrow 0$ thus the exponent of the provided upper bound matches the optimal exponent in \eqref{eq:exp_id}.

\vspace{5 pt}
\textit{iii)} In the regime $\epsilon {|\mathcal X|} \ll 1 $, we have  $\log(1+|{ \mathcal X}|\epsilon) \approx 0$ and  $\bar{P}_j= (Q_j(x)+\epsilon_j)/(1+|{\mathcal X}|\epsilon_j) \approx Q_j(x)$. Therefore, the exponent in the upper bound \eqref{eq:exp_bound} approximately equals to
\begin{align}
\text{min}_{i \neq j}  C(Q_i,Q_j),  \quad i=1,2,...,M, \quad j=1,2,...,M. 
\end{align}
This is the minimum pairwise Chernoff information between the distinct nominal distributions. Thus, in this regime, the nominal distributions act as if they are the true distributions when
the test is of the form \eqref{eq:prop_test}.  We have  
\begin{align}
\text{min}_{i \neq j}  C(Q_i,Q_j) &\geq \text{min}_{i \neq j}  -\frac{1}{2} \ln (1-V(Q_i,Q_j)^2) \\
                                                   & \geq \text{min}_{i \neq j}  \frac{1}{2}V(Q_i,Q_j)^2 \label{eq:C_bound}
\end{align}
where the first inequality results from \eqref{eq:Sason_ineq} and the second one is due to the inequality $ \ln(z) \leq z-1$, $z \geq 0$. 
The right hand side of \eqref{eq:C_bound} is the exponent of the upper bound of the DGL test in \cite{Devroye} when $P_j =Q_j$, $j=1,2,....,M$, i.e. when there is no uncertainty in the hypothesis distributions. Therefore, in the regime $\epsilon {|\mathcal X|} \ll 1$, the achievable exponent of the proposed method is larger than the exponent of the error probability upper bound of DGL test \footnote{The proposed method is only applicable when $|\mathcal X|$ is finite, however the DGL test can also be applied when the underlying alphabet is continuous i.e. $|\mathcal X|=\infty$.}. This observation is also validated via simulations as we demonstrate in the next section.

\begin{figure*}[t]
\begin{center}
\includegraphics[scale=0.22]{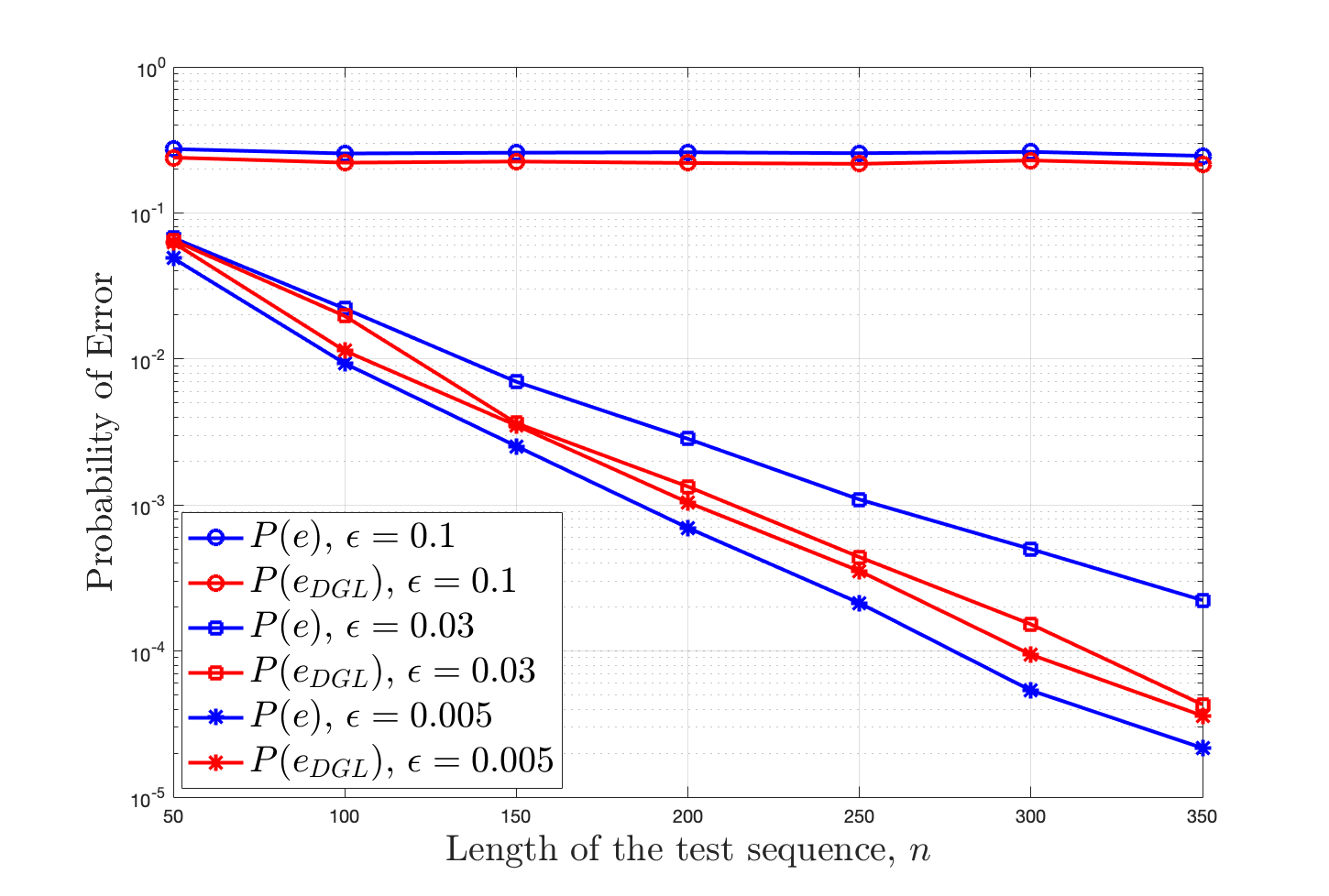}
\caption{Performance comparison of the DGL test and the proposed method.}
\label{fig: fig_2}
\end{center}
\end{figure*}

\vspace{5 pt}

%
%
 \begin{table*}[t]
\caption{The nominal distributions used for the simulations demonstrated in Figure 2}
\begin{center}
\resizebox{ 1 \textwidth}{!}
{
\begin{tabular}{|c|c|c|c|c|c|c|c|c|}
\hline
  & \textbf{$Q_1$} & \textbf{$Q_2$} & \textbf{$Q_3$} & \textbf{$Q_4$} & \textbf{$Q_5$} 
\\
\hline
 $\epsilon=0.1$ &$[0.04,0.76,0.2]$ & $[0.24,0.3,0.46];$&$[0.7,0.05,0.25]$&$[0.37,0.5,0.13]$&$[0.34,0.5,0.16]$
\\
\hline
 $\epsilon=0.03$ & $[0.11,0.82,0.07]$&$[0.29,0.23,0.48]$&$[0.63,0.09,0.28]$&$[0.38,0.43,0.19]$&$[0.32,0.57,0.11]$ 
\\
\hline
 $\epsilon=0.005$ & $[0.102,0.803,0.095]$&$[0.305,0.198,0.497]$&$[0.599,0.096,0.305]$&$[0.398,0.397,0.205]$&$[0.305,0.599,0.096]$
\\
\hline

\end{tabular}
}
\end{center}
\label{table:table_1}
\end{table*}

\subsection{Simulations} 

First, we have performed simulations to compare the performance of the DGL test and the proposed method. We have considered the case $M=5$ and $|{\mathcal X}|=3$ where the distributions of the hypothesis are  
\begin{align*}
&P_1= [0.1,0.8,0.1], \quad
P_2= [0.3,0.2,0.5], \quad
P_3= [0.6,0.1,0.3], \quad \\
& \quad \quad \quad \quad P_4= [0.4,0.4,0.2], \quad
P_5= [0.3,0.6,0.1],
\end{align*}
and the nominal distributions are listed in Table 1.

The simulated error probabilities are demonstrated in Figure \ref{fig: fig_2} where $P(e)$ and $P(e_{DGL})$ denote the error probability of the proposed method and the DGL test,
respectively. We observe that when $\epsilon=0.1$ consistent detection is not possible with both methods. This results from the fact that the minimum separation between the source distributions is not sufficient for both tests. As $\epsilon$ decreases both tests start to perform consistently, and when $\epsilon=0.03$ the DGL test performs better than the proposed method. However, as $\epsilon$ further decreases the performance of the proposed methods catches up with that of DGL test and when $\epsilon=0.005$ it performs better. The same trend holds as $\epsilon \rightarrow 0$, as well. Therefore, the proposed method seems to provide a performance advantage for arbitrarily small values of $\epsilon$ as suggested by implication \textit{iii)} of Theorem  \ref{thm:thm_1}. A detailed  performance comparison of the proposed method and the DGL test can be found in \cite{Yildirim}.

Next, we have considered the case where the hypothesis distributions are not represented exactly due to hardware limitations. This may happen if the system where the test is being performed has limited memory or computational resources \cite{Varshney}. If this is the case, one has to quantize the hypothesis distributions with a precision that is allowed by the system and perform the test with the resulting nominal distributions. Here, we give consider such an application where the hypothesis distribution vectors are quantized into $q$, $q=1,2,....,$ bits. Let $p$, $ p \in [0,1]$, be a real number representing an element of the distribution vector and $z$ be an unsigned integer represented with $q$ bits. In the assumed quantization model (Matlab's fixed point design model with slope and bias representation) $p$ is represented by $\hat{p}$, $\hat{p} \in [0,1]$, as 
\begin{align}
\hat{p} = \text{integer value of } z \times 2^{-q}+ \text{bias}
\end{align}
and $z$ is selected such that the difference between $p$ and $\hat{p}$ is smallest. We have applied the above quantization model to hypothesis distributions by setting the bias to 0. In order to ensure that the quantized versions are actual distributions i.e. their elements add up to 1, we have applied quantization to first $|{\mathcal X}|-1$ elements of distribution vectors, then the last element is obtained by subtracting the sum of quantized versions from 1. The resulting nominal distributions, $Q_1,Q_2,...,Q_M$, and the parameters are listed in Table \ref{table:table_1}.

\begin{figure*}[t]
\begin{center}
\includegraphics[scale=0.22]{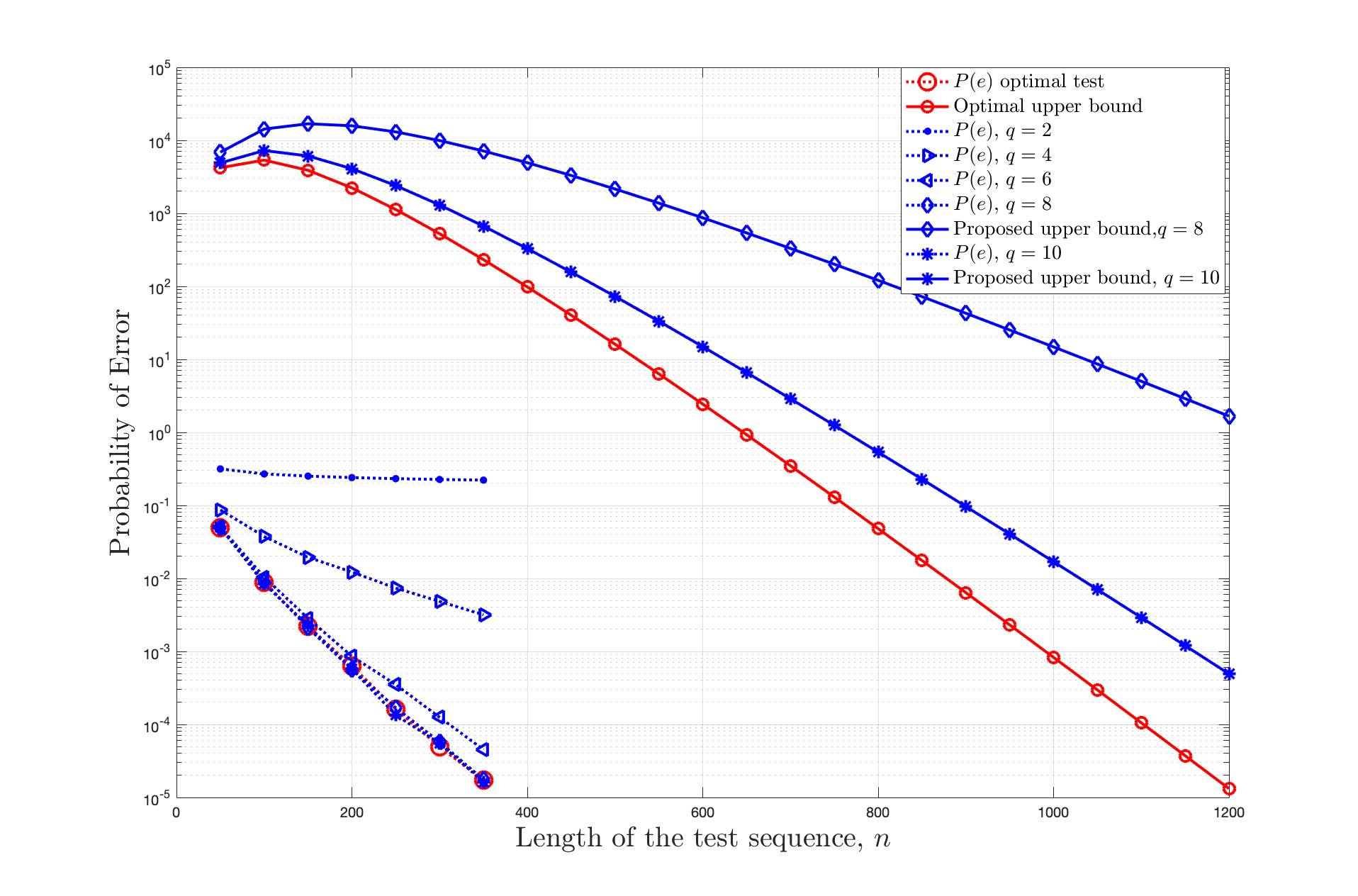}
\caption{ Simulated $P(e)$ values and theoretical upper bounds, where $q$ denotes the number of bits used for representing the nominal distributions.}
\label{fig:fig_3}
\end{center}
\end{figure*}

The simulated error probabilities, $P(e)$, along with the proposed upper bound \eqref{eq:exp_bound} are presented in Figure \ref{fig:fig_3}, where we have also plotted the performance of the optimal test  \eqref{eq:NN_rule} and the optimal upper bound \eqref{eq:exp_id} for comparison. We have gradually increased the number of bits, $q$, used in quantization and observed the performance of the proposed method.  In the simulations, the exponents are the negative slopes of the $P(e)$ curves and the upper bounds, and as evident from Figure \ref{fig:fig_3} these are in agreement with the proposed analysis. For the considered setup $\text{min}_{i \neq j}  C(P_i,P_j)=0.0329$ and the proposed test provides an acceptable performance for 
$q \geq 4$. When $q=2,4,6$ we observe that $C(\bar{P}_i,\bar{P}_j) < \log(1+ |\mathcal X| \epsilon)$ thus the exponent term in \eqref{eq:exp_bound} is negative and the provided upper bound does not guarantee an achievable exponent. However, $P(e)$ curves have negative slopes which implies an acceptable performance. For the case $q=8,10$, $C(\bar{P}_i,\bar{P}_j) > \log(1+ |\mathcal X| \epsilon)$  thus the proposed upper bound is active and the performance of the proposed test closely matches the optimal test.

 \begin{table*}[t]
\caption{The nominal distributions for the simulations demonstrated in Figure 3}
\begin{center}
\resizebox{ 1 \textwidth}{!}
{
\begin{tabular}{|c|c|c|c|c|c|c|c|c|}
\hline
  & \textbf{$Q_1$} & \textbf{$Q_2$} & \textbf{$Q_3$} & \textbf{$Q_4$} & \textbf{$Q_5$} & \textbf{$ \underset{i \neq j}{\textnormal{min }} C(\bar{P}_i,\bar{P}_j)$} & 
 \textbf{$\log(1+ |\mathcal X| \epsilon)$}
\\
\hline
 $q=2$ &$[0,0.7500,0.2500]$ & $[0.2500,0.2500,0.5000]$&$[0.5000,0,0.5000]$&$[0.5000,0.5000,0]$&$[0.2500,0.5000,0.2500]$ & $0.0351$ & $0.6781$
\\
\hline
 $q=4$ & $[0.1250,0.8125,0.0625]$&$[0.3125,0.1875,0.5000]$&$[0.6250,0.1250,0.2500]$&$[0.3750,0.3750,0.2500]$&$[0.3125,0.6250,0.0625]$ & $0.0310$ & $0.2016$
\\
\hline
 $q=6$ & $[0.0938,0.7969,0.1094]$&$[0.2969,0.2031,0.5000]$&$[0.5938,0.0938,0.3125]$&$[0.4062,0.4062,0.1875]$&$[0.2969,0.5938,0.1094]$ & $0.0186$ & $0.0531$
\\
\hline
 $q=8$ & $[0.1016,0.8008,0.0977]$&$[0.3008,0.1992,0.5000]$&$[0.6016,0.1016,0.2969]$&$[0.3984,0.3984,0.2031]$&$ 0.3008,0.6016,0.09777]$ & $0.0318$ & $0.00135$
\\
\hline
 $q=10$ & $[0.0996,0.7998,0.1006]$&$[0.2998,0.2002,0.5000]$&$[0.5996,0.0996,0.3008]$&$[0.4004,0.4004,0.1992]$&$[0.2998,0.5996,0.1006]$ & $0.0315$ & $0.0034$
\\
\hline

\hline
\end{tabular}
}
\end{center}
\label{table:table_1}
\end{table*}

\section{Concluding remarks} \label{section:Sect_5}

In this paper, we have investigated Bayesian MHT by providing a type-based analysis on its error probability. We have shown that the worst case achievable
error exponent, over the distinct types, equals to the minimum Chernoff information between the distinct pairs of hypothesis distributions. The proposed analysis extends Sanov's 
theorem-based methods by providing insight on the error probabilities of specific input types.

As a generalization of the proposed analysis, we have considered  the case where the true distributions of the hypothesis are not known exactly, but a set of nominal distributions that are close to the true distributions is available. We have proposed a simple type-based test, obtained an upper bound on its error probability and showed that it coincides with the optimal NN decision rule as the uncertainties in distributions vanish. We have compared its performance with the robust DGL test and show that the achievable error exponent of the proposed method is larger when the nominal distributions are arbitrarily close to true ones. This observation is also validated with simulations. Being identical in form to the NN test, the proposed method has
complexity $O( M n)$ which is linear in the number of hypothesis and in the length of the test sequence. Thus, it also provides a complexity advantage over DGL test whose complexity is 
$O( M^2 n + M^2 \log M)$ \cite{Devroye}. This advantage is particularly important when the number of hypothesis, $M$, is large. One drawback of the proposed method is that, being a type-based method, it is applicable only when hypothesis distributions are discrete, whereas the DGL test can be used with continuous distributions, as well.

In this work, we have assumed that the nominal distributions are close to true distributions in variational distance, but our method is also applicable if these distances are bounded in terms of $\ell_1$, separation, Hellinger, Wasserstein, $\chi^2$ or KL distance. Because any of these upper bounds implies another upper bound on the variational distance \cite{Gibbs}.  In this regard, KL divergence plays an important role if the nominal distributions are estimated from labeled training sequences. Let $\vec{t}_j=[t_{j1},t_{j2},...,t_{jm}]$ be a training sequence that is generated from distribution $P_j$, and let $P_{\vec{t}_j}$ denote the type of $\vec{t}_j$. The law of large numbers states that \cite[p. 356]{Cover}
\begin{align}
 \Pr \left( D( P_{\vec{t}_j}|| P_j \right) > \beta ) \leq 2^{-m\left( \beta-\frac{|{\mathcal X}|\log(m+1)}{m} \right)}
\end{align} 
Thus, $P_{\vec{t}_j}$ can be used as a nominal distributions for $P_j$. Then, by changing $\beta$ one can adjust the estimation accuracy, and by changing $m$
one can adjust the resulting error probability of the estimation. In turn, one can analyze the performance of multiple classification with labeled training sequences.
Recently, we have followed this approach and investigated the classification performance of the DGL test and obtained simple upper bounds on its error probability \cite{Afser}. 
Investigating the classification performance of the  proposed method in this paper is the topic of our upcoming work.

\section{APPENDIX}
\vspace{10pt}
\subsection{Proof of Proposition \ref{thm:prop_1}}
\vspace{5pt}

For fixed $i$ and $j$,  $i \neq j$, let $P(e_j |{\mathcal H}_i, P_{\vec{x}} )$ denote the probability of error, averaged over ${\vec{x}} \in {\mathcal{T}}(P_{\vec{x}})$, when ${\mathcal H}_i$ is true but the test decides on ${\mathcal H}_j$. This corresponds to the event that $D(P_{{\vec{x}}}||P_j) < \underset{k \neq j}{\min} \hspace{2 pt} D(P_{{\vec{x}}}||P_k)$, $k=1,2,...,M$, when ${\vec{x}}$
is generated from distribution $P_i$. Thus,
\begin{align}
P(e_j |H_i,P_{\vec{x}} ) &= \sum\limits_{\substack{{\vec{x}} \in  T(P_{\vec{x}}) \\ D(P_{{\vec{x}}}||P_j) < \underset{k \neq j}{\min} \hspace{2 pt} D(P_{{\vec{x}}}||P_k)}} P_i({\vec{x}}), \quad k=1,2,...,M \nonumber 
\\ & {=} \hspace{-10pt}\sum\limits_{\substack{{\vec{x}} \in  T(P_{\vec{x}})) \\ D(P_{{\vec{x}}}||P_j) < \underset{k \neq j}{\min} \hspace{2 pt} D(P_{{\vec{x}}}||P_k)}} 2^{-n(D(P_{{\vec{x}}}||P_i)+H(P_{{\vec{x}}}))},  \quad k=1,2,...,M  \nonumber \\
& {\leq} \hspace{-10pt}\sum\limits_{\substack{{\vec{x}} \in  T(P_{\vec{x}}) \\ D(P_{{\vec{x}}}||P_j) < D(P_{{\vec{x}}}||P_i) }} 2^{-n(D(P_{{\vec{x}}}||P_i)+H(P_{{\vec{x}}}))}  \label{eq:bound_a}
\end{align}
where the inequality follows since the event  $D(P_{{\vec{x}}}||P_j) < \underset{k \neq j}{\min} \hspace{2 pt} D(P_{{\vec{x}}}||P_k)$ is a subset of $D(P_{{\vec{x}}}||P_j) < D(P_{{\vec{x}}}||P_i)$. Interchanging the roles of  $i$ and $j$ we obtain
\begin{align}
P(e_i |{\mathcal H}_j, P_{\vec{x}} ) \leq \hspace{-10pt}\sum\limits_{\substack{{\vec{x}} \in T(P_{\vec{x}}) \\ D(P_{{\vec{x}}}||P_i) < D(P_{{\vec{x}}}||P_j) }} 2^{-n(D(P_{{\vec{x}}}||P_j)+H(P_{{\vec{x}}}))}  \label{eq:bound_b}
\end{align}
We combine \eqref{eq:bound_a} to  \eqref{eq:bound_b} to upper bound $P(e_j |H_i, T(P_{\vec{x}} )) $ as

\begin{align}
P(e_j |{\mathcal H}_i, P_{\vec{x}} )  &\leq P(e_j |{\mathcal H}_i,  P_{\vec{x}} ) + P(e_i |{\mathcal H}_j,P_{\vec{x}}) \nonumber \\
		   &\leq \sum\limits_{\substack{{\vec{x}} \in  {\mathcal T}(P_{\vec{x}}) \\ D(P_{{\vec{x}}}||P_j) < D(P_{{\vec{x}}}||P_i) }} 2^{-n(D(P_{{\vec{x}}}||P_i)+H(P_{{\vec{x}}}))}   +
		  \hspace{-10 pt} \sum\limits_{\substack{{\vec{x}} \in  {\mathcal T}(P_{\vec{x}}) \\ D(P_{{\vec{x}}}||P_i) < D(P_{{\vec{x}}}||P_j) }} 2^{-n(D(P_{{\vec{x}}}||P_j)+H(P_{{\vec{x}}}))}  \nonumber  \\
	&=	   \sum_{{\vec{x}} \in {\mathcal T}(P_{\vec{x}}) } \hspace{-5pt} 2^{-n \left(  \text{max} \{ D(P_{{\vec{x}}}||P_i), D(P_{{\vec{x}}}||P_j) \} +H(P_{{\vec{x}}}) \right)} \nonumber  \\
	&= | T(P_{\vec{x}}) | 2^{-n \left(  \text{max} \{ D(P_{{\vec{x}}}||P_i), D(P_{{\vec{x}}}||P_j) \}+H(P_{{\vec{x}}}) \right)} \nonumber  \\
	&\leq 2^{-n \left(  \text{max} \{ D(P_{{\vec{x}}}||P_i), D(P_{{\vec{x}}}||P_j) \} \right)}  \label{eq:upper_bound_1}
\end{align}
where the last inequality follows from Lemma \ref{thm:lemma_1}. The error probability when $H_i$ is true is
\begin{align}
P(e|{\mathcal H}_i, P_{\vec{x}}) &= \sum_{j \neq i, }  P(e_j |{\mathcal H}_i, P_{\vec{x}}), \quad j=1,2,...,M  \nonumber \\
 &\leq M \underset{j \neq i}{\max} P(e_j |{\mathcal H}_i, P_{\vec{x}}), \quad j=1,2,...,M  \nonumber \\
 &\leq M \hspace{1pt} \underset{j \neq i}{\max} \hspace{1pt} 2^{-n \left(  \text{max} \{ D(P_{{\vec{x}}}||P_i), D(P_{{\vec{x}}}||P_j) \} \right)},  \quad j=1,2,...,M \nonumber  \\
 &=   2^{-n \left(  \underset{j \neq i}{\min} \hspace{3pt} \text{max} \{ D(P_{{\vec{x}}}||P_i), D(P_{{\vec{x}}}||P_j) \} - \frac{\log M}{n} \right)},  \quad j=1,2,...,M.
\end{align}
Finally, $P(e| P_{\vec{x}} )$ can be upper bounded as 
\begin{align}
P(e| P_{\vec{x}} )  &= \sum_{i=1}^{M}P(e |{\mathcal H}_i,  P_{\vec{x}} ) P({\mathcal H}_i) \nonumber \\
&\leq \underset{i}{\max} P(e |{\mathcal H}_i,  P_{\vec{x}} ),  \quad i=1,2,....,M  \nonumber \\
&\leq  \underset{i}{\max}  2^{-n \left( \underset{j \neq i}{\min} \hspace{3pt} \text{max} \{ D(P_{{\vec{x}}}||P_i), D(P_{{\vec{x}}}||P_j) \} - \frac{\log M}{n} \right)}, \quad i,j=1,2,...,M  \nonumber      \\
           &=  2^{-n \left(  \underset{i \neq j}{\min} \hspace{3pt} \text{max} \{ D(P_{{\vec{x}}}||P_i), D(P_{{\vec{x}}}||P_j) \} - \frac{\log M}{n} \right)}, \quad i,j=1,2,...,M
\end{align}
which completes the proof.

\subsection{Proof of Proposition \ref{thm:prop_2}}
We are interested in finding the type class $T(P_{\vec{x}}) \in {\mathcal P}^n$ that minimizes 
$\text{max} \{ D(P_{{\vec{x}}}||P_i), D(P_{{\vec{x}}}||P_j) \}$. This search can be transformed into a constrained optimization problem as
\begin{align}
&\underset{T(P_{\vec{x}}) \in {\mathcal P}^n}{\text{minimize}} \quad D(P_{{\vec{x}}}||P_i), \nonumber \\
&\text{subject to} \quad D(P_{{\vec{x}}}||P_i)-D(P_{{\vec{x}}}||P_j) \geq 0. \label{eq:optimization}
\end{align} 
Using the method of Lagrange multipliers we obtain
\begin{align*}
J(P_{\vec{x}}) &= \sum_{x \in P_{\vec{x}}} P_{\vec{x}}(x) \log \frac{P_{\vec{x}}(x)}{\bar{P}_i(x)} + \lambda  \sum_{x \in P_{\vec{x}}} P_{\vec{x}}(x)\log \frac{P_i(x)}{P_j(x)}  + v \sum_{x \in P_{\vec{x}}} P_{\vec{x}}(x) 
\end{align*}
where $\lambda$ and $v$ are constants. Differentiating with respect to $P_{\vec{x}}$ yields
\begin{align*}
\log \frac{P_{\vec{x}}(x)}{\bar{P}_i(x)} + 1+\lambda \log \frac{P_i(x)}{P_j(x)} + v=0.
\end{align*}
Solving the above equation reveals that the minimizer $P_{\vec{x}}$ must be of the form
\begin{align}
P_{{\vec{x}}}^\lambda= \frac{P_i(x)^\lambda P_j(x)^{1-\lambda}}{\sum_{x \in {\mathcal X}} P_i(x)^\lambda P_j(x)^{1-\lambda} }
\label{eq:p_form}
\end{align}
and $\lambda$ is chosen such that $D(P_{{\vec{x}}}^\lambda | \bar{P}_i) = D(P_{{\vec{x}}}^\lambda | \bar{P}_j)$. When $P_{\vec{x}}$ has the form in \eqref{eq:p_form}, the condition $D(P_{{\vec{x}}}^\lambda |P_i) = D(P_{{\vec{x}}}^\lambda | P_j)$ is equivalent to the definition of the Chernoff distance and $D(P_{{\vec{x}}}^\lambda | P_i) = D(P_{{\vec{x}}}^\lambda | P_j)= C( \bar{P}_i, \bar{P}_j)$ holds (see \cite[p. 385]{Cover}). Besides, interchanging the roles of $D(P_{{\vec{x}}}||\bar{P}_i)$ and $D(P_{{\vec{x}}}||\bar{P}_j)$ in the constraint optimization problem \eqref{eq:optimization} ends up with the same result. Therefore, we conclude that 
\begin{align}
 \underset{ T(P_{\vec{x}}) \in {\mathcal P}^n}{\text{min}}  \text{max} \{ D(P_{{\vec{x}}}||P_i), D(P_{{\vec{x}}}||P_j) \}  = C(P_i, P_j). \label{eq:Cher_eq}
\end{align}

\subsection{Proof of Proposition \ref{thm:prop_3}}

\begin{align*}
P_j({\vec{x}}) &= \prod_{i=1}^n P_j(x_i)
\\& =\prod_{a\in \mathcal{X}} P_j(a)^{n P_{{\vec{x}}}(a)}
\\&\leq \prod_{a\in \mathcal{X}} {(Q_j(a)+\epsilon_j)}^{n P_{{\vec{x}}}(a)}
\\&=\prod_{a\in \mathcal{X}} 2^{n P_{{\vec{x}}}(a)\log({Q_j(a)+\epsilon_j})}
\\&=\prod_{a\in \mathcal{X}} 2^{n P_{{\vec{x}}}(a)\log{\frac{(Q_j(a)+\epsilon_j)(1+|{ \mathcal X}|\epsilon_j) P_{{\vec{x}}}(a)}{(1+|{ \mathcal X}|\epsilon_j )P_{{\vec{x}}}(a)}}}
\\&=2^{-n\sum_{a\in \mathcal{X}}P_{{\vec{x}}}(a)\left( \log{\frac{P_{{\vec{x}}}(a)}{\frac{Q_j(a)+\epsilon_j}{1+|{ \mathcal X}|\epsilon_j}}}+\log{\frac{1}{P_{{\vec{x}}}(a)}+\log{\frac{1}{1+|{ \mathcal X}|\epsilon_j}}}\right)}
\\&=2^{-n\sum_{a\in \mathcal{X}}P_{{\vec{x}}}(a)\left( \log{\frac{P_{{\vec{x}}}(a)}{\bar{P}_j(a)}}+\log{\frac{1}{P_{{\vec{x}}}(a)}+\log{\frac{1}{1+|{ \mathcal X}|\epsilon_j} } } \right) }
\\&=2^{-n\left( D(P_{{\vec{x}}}||\bar{P}_j)+H(P_{{\vec{x}}})-\log({1+|{ \mathcal X}|\epsilon_j}) \right)}
\end{align*}
where the inequality follows from the variational distance constraint i.e. $\frac{1}{2}\sum_{x \in \mathcal{X}} |P_j(a)-Q_j(a)| \leq \epsilon_j$ and $\sum_{x \in \mathcal{X}}Q_j(x)=1$.

\end{document}